# Galilean invariance of Hamiltonian guiding center theory

C. D. Stephens[a] and J. W. Burby

AFFILIATIONS

Institute of Fusion Studies, University of Texas at Austin, Austin, Texas 78712-1192, USA

[a]Author to whom correspondence should be addressed: cstephens@realtafusion.com

## ABSTRACT

Galilean invariance is a desired property for classical non-relativistic theories. Guiding center theory describes the slow drift motion of charged particles in strong magnetic fields. While one can guarantee a Hamiltonian formulation to all orders, it was previously unknown whether such a theory can possess Galilean symmetry. We show that specific first- and second-order guiding center models are Galilean invariant. Furthermore, we prove that one can ensure Galilean invariance to all orders.

## I. INTRODUCTION

The principle of relativity states that the laws of physics are identical in all inertial reference frames. A classical, non-relativistic theory that respects this principle is form-invariant with respect to Galilean boosts. When such a theory is also form-invariant with respect to translations and rotations, we say the theory possesses Galilean symmetry and is called Galilean invariant. Quantities that do not change under Galilean transformations are called Galilean invariants, and equations of motion that are form-invariant are called Galilean covariant. Two fundamental Hamiltonian kinetic models of fully ionized plasmas are Galilean invariant. The first is Vlasov–Poisson (VP),[1] which is given by

$$\frac{\partial f_s}{\partial t} + \mathbf{v}\cdot\nabla f_s + \frac{e_s}{m_s}\mathbf{E}\cdot\frac{\partial f_s}{\partial \mathbf{v}} = 0, \tag{1}$$

$$\nabla\cdot\mathbf{E} = \frac{1}{\epsilon_0}\sum_s \int \mathrm{d}^3 v\, e_s f_s, \tag{2}$$

$$\nabla\times\mathbf{E} = 0. \tag{3}$$

Here, $s$ signifies the species, $f$ is the distribution function defined on the phase space $(\mathbf{x}, \mathbf{v})$ and time $t$, $e$ being the charge, $m$ being the mass, $\mathbf{E}$ being the electric field, and $\epsilon_0$ being the electric constant. In VP, $f$ and $\mathbf{E}$ are both Galilean invariant quantities. The second theory is quasineutral Vlasov (QV),[2] which is given by

$$\frac{\partial f_s}{\partial t} + \mathbf{v}\cdot\nabla f_s + \frac{e_s}{m_s}(\mathbf{E} + \mathbf{v}\times\mathbf{B})\cdot\frac{\partial f_s}{\partial \mathbf{v}} = 0, \tag{4}$$

$$0 = \sum_s \int \mathrm{d}^3 v\, e_s f_s, \tag{5}$$

$$\nabla\cdot\mathbf{B} = 0, \tag{6}$$

$$\nabla\times\mathbf{E} = -\frac{\partial \mathbf{B}}{\partial t}, \tag{7}$$

$$\nabla\times\mathbf{B} = \mu_0 \sum_s \int \mathrm{d}^3 v\, e_s \mathbf{v} f_s, \tag{8}$$

where the magnetic field $\mathbf{B}$ and magnetic constant $\mu_0$ are introduced. In QV, while $f$ and $\mathbf{B}$ are Galilean invariant quantities, $\mathbf{E}$ transforms as $\mathbf{E}\to\mathbf{E}' = \mathbf{E} + \mathbf{u}\times\mathbf{B}$, where $\mathbf{u}$ is the constant velocity of the primed frame $S'$ relative to the unprimed frame $S$. This ensures that the Lorentz force is Galilean invariant, as $\mathbf{v}\to\mathbf{v}' = \mathbf{v} - \mathbf{u}$. Both VP and QV arise as distinct limits of the fully relativistic Vlasov–Maxwell system. Galilean invariance requires that the electromagnetic force density $\mathbf{f} = \rho\mathbf{E} + \mathbf{J}\times\mathbf{B}$ on the conducting fluid is Galilean invariant, where $\rho$ is the total charge density and $\mathbf{J}$ is the total current.[3,4] In VP, the absence of a magnetic field means $\mathbf{f} = \rho\mathbf{E}$, while quasineutrality in QV leads to $\mathbf{f} = \mathbf{J}\times\mathbf{B}$.[5]

In strongly magnetized plasmas where the gyroradius is smaller than the relevant macroscopic length scale, the fast cyclotron dynamics can be eliminated from QV. In doing so, one decomposes the particle trajectory into a fast periodic orbit and a slowly moving guiding center, and the resulting equations of motion for the guiding center constitute the guiding center theory.[6,7] Furthermore, the magnetic moment $\mu$ is an adiabatic invariant associated with the fast periodic motion[8,9] and is thus conserved in the guiding center equations of motion. Lie transform perturbation theory[10] can be employed to ensure that the model is Hamiltonian to all orders and gyrophase-invariant up to arbitrary order.

Since the non-relativistic Lorentz force equations possess Galilean symmetry, it is reasonable to ask if one can formulate a guiding center theory that preserves the symmetry. Moreover, relativistic theories of guiding center motion that respect Lorentz symmetry can be found in the literature.[11–13] Since the Lorentz invariance is guaranteed by the covariant structure of relativistic guiding center theory, one should expect to recover Galilean invariance to all orders upon taking the non-relativistic limit. Rather than proceeding upon those lines, we take the non-relativistic limit first and then analyze Galilean symmetry directly. Not only is the mathematical machinery easier to digest in the non-relativistic limit, but it also lends itself to the explicit construction of a Galilean invariant Hamiltonian guiding center theory to any desired order. Constructing a Lorentz invariant theory up to a specified order and then taking the non-relativistic limit should, of course, lead to an equivalent result.

For the purpose of this article, we specifically analyze form-invariance under Galilean boosts, as form-invariance under translations and rotations is uncontroversial.[6] As far as these authors are aware, only a handful of publications have considered Galilean invariance of the guiding center model.[14–16] Although Ref. 16 correctly recognizes a definition of the magnetic moment that respects Galilean invariance, which itself dates to the 1960s,[17] its discussion of Galilean invariance of the guiding center model itself is limited. Although any reasonable Galilean invariant guiding center model must include a Galilean invariant definition of the magnetic moment, a detailed analysis of the underlying model is still required to demonstrate Galilean invariance. Meanwhile, Ref. 15 conjectures that there exists no Hamiltonian guiding center theory that respects the Galilean invariance. Otherwise, the topic is conspicuously absent from the literature. In this article, we fill this gap by presenting Hamiltonian guiding center theories that are indeed Galilean invariant. We also rigorously show that the Galilean invariance can be preserved to all orders.

This article is structured as follows: in Sec. II, we examine the first-order guiding center equations of motion and show that they are Galilean covariant (i.e., form-invariant). In Sec. III, we show that the single-particle second-order guiding center action possesses Galilean divergence symmetry, showing that the second-order equations of motion are also Galilean covariant. We also show that the resulting QV theory must therefore also be Galilean invariant, whereby the guiding center polarization and magnetization fields transform as expected under a Galilean boost. In Sec. IV, we prove that one can construct a Hamiltonian guiding center theory to arbitrary order. We discuss general requirements for properly constructing a Galilean invariant theory in Sec. V, and why certain common forms of Hamiltonian guiding center theories are not Galilean invariant. Finally, we summarize this work in Sec. VI.

## II. GALILEAN COVARIANCE OF THE FIRST-ORDER GUIDING CENTER EQUATIONS OF MOTION

We write the first order guiding center Lagrangian and Hamiltonian in the strong flow regime as[7,18–20]

$$L = \left(e\mathbf{A} + mv_\parallel\hat{\mathbf{b}} + m\mathbf{u}_E\right)\cdot\frac{\mathrm{d}\mathbf{X}}{\mathrm{d}t} + \frac{m\mu}{e}\frac{\mathrm{d}\alpha}{\mathrm{d}t} - H, \tag{9}$$

$$H = e\phi + \frac{m}{2}\left(v_\parallel\hat{\mathbf{b}} + \mathbf{u}_E\right)^2 + \mu B. \tag{10}$$

Here, $\phi$ and $\mathbf{A}$ are the electromagnetic potentials, $\mathbf{X}$ is the guiding center position, $v_\parallel$ is the parallel velocity, $\hat{\mathbf{b}} = \mathbf{B}/B$ is the magnetic field direction, $\mathbf{u}_E = \mathbf{E}/B\times\hat{\mathbf{b}}$ is the $E$-cross-$B$ velocity, and $\alpha$ is the gyrophase. From the Euler–Lagrange equations,

$$\frac{\mathrm{d}}{\mathrm{d}t}\frac{\partial L}{\partial\dot{Z}^i} = \frac{\partial L}{\partial Z^i}, \tag{11}$$

we can derive the equations of motion as

$$m\frac{\mathrm{d}v_\parallel}{\mathrm{d}t}\hat{\mathbf{b}} = e\mathbf{E}^* + e\dot{\mathbf{X}}\times\mathbf{B}^* - \mu\nabla B - \frac{m}{2}\nabla|\mathbf{u}_E|^2, \tag{12}$$

$$\hat{\mathbf{b}}\cdot\frac{\mathrm{d}\mathbf{X}}{\mathrm{d}t} = v_\parallel, \tag{13}$$

$$\dot{\alpha} = \frac{eB}{m}, \tag{14}$$

$$\dot{\mu} = 0, \tag{15}$$

where

$$\mathbf{A}^* = \mathbf{A} + \frac{m}{e}\left(v_\parallel\hat{\mathbf{b}} + \mathbf{u}_E\right), \tag{16}$$

$$\mathbf{E}^* = -\nabla\phi - \frac{\partial\mathbf{A}^*}{\partial t}, \tag{17}$$

$$\mathbf{B}^* = \nabla\times\mathbf{A}^*. \tag{18}$$

The equations of motion for the guiding center and the parallel acceleration are then typically expressed in the following form:

$$\frac{\mathrm{d}\mathbf{X}}{\mathrm{d}t} = \frac{v_\parallel\mathbf{B}^*}{B_\parallel^*} + \frac{m\hat{\mathbf{b}}}{eB_\parallel^*}\times\left(\frac{\mu}{m}\nabla B - \frac{e}{m}\mathbf{E}^* + \frac{1}{2}\nabla|\mathbf{u}_E|^2\right), \tag{19}$$

$$\frac{\mathrm{d}v_\parallel}{\mathrm{d}t} = -\frac{\mathbf{B}^*}{B_\parallel^*}\cdot\left(\frac{\mu}{m}\nabla B - \frac{e}{m}\mathbf{E}^* + \frac{1}{2}\nabla|\mathbf{u}_E|^2\right), \tag{20}$$

where $B_\parallel^* = \mathbf{B}^*\cdot\hat{\mathbf{b}}$. However, these equations are inconvenient for proving Galilean covariance and would involve several tedious calculations because the effective fields $\mathbf{E}^*$ and $\mathbf{B}^*$ do not transform in a straightforward way.

Instead, we write the equations of motion

$$\frac{\mathrm{d}\mathbf{U}_0}{\mathrm{d}t} = \frac{e}{m}\mathbf{E} + \frac{e}{m}\frac{\mathrm{d}\mathbf{X}}{\mathrm{d}t}\times\mathbf{B} - \frac{\mu}{m}\nabla B + (\nabla\mathbf{U}_0)\cdot\mathbf{U}_1, \tag{21}$$

$$\hat{\mathbf{b}}\cdot\frac{\mathrm{d}\mathbf{X}}{\mathrm{d}t} = v_\parallel, \tag{22}$$

where $(\nabla\mathbf{A})\cdot\mathbf{B} = \mathbf{B}\cdot\nabla\mathbf{A} + \mathbf{B}\times(\nabla\times\mathbf{A})$. They can be condensed into the following equations:

$$\frac{\mathrm{d}\mathbf{X}}{\mathrm{d}t} = \mathbf{U}_0 + \mathbf{U}_1, \tag{23}$$

$$\hat{\mathbf{b}}\cdot\frac{\mathrm{d}\mathbf{U}_0}{\mathrm{d}t} = \frac{e}{m}\hat{\mathbf{b}}\cdot\mathbf{E} - \frac{\mu}{m}\hat{\mathbf{b}}\cdot\nabla B + \mathbf{U}_1\cdot\left(\hat{\mathbf{b}}\cdot\nabla\mathbf{U}_0\right), \tag{24}$$

where

$$\mathbf{U}_0 = v_\parallel\hat{\mathbf{b}} + \mathbf{u}_E, \tag{25}$$

$$\mathbf{U}_1 = \frac{m\hat{\mathbf{b}}}{eB_\parallel^*} \times \left(\frac{\mu}{m}\nabla B + \frac{\mathrm{D}\mathbf{U}_0}{\mathrm{D}t}\right), \tag{26}$$

$$\frac{\mathrm{D}}{\mathrm{D}t} = \frac{\partial}{\partial t} + \mathbf{U}_0 \cdot \nabla. \tag{27}$$

We see that the guiding center velocity consists of parallel streaming motion, the classical $E$-cross-$B$ drift, a modified grad-$B$ drift, and a modified inertial drift. The modified drifts differ from the classical drifts due to the presence of $B_\parallel^*$ and therefore include curvature drift effects.

The Galilean boost can be defined with the following transformations:

$$\mathbf{X}' = \mathbf{X} - \mathbf{u}t, \qquad t' = t, \tag{28}$$

$$v_\parallel' = v_\parallel - \mathbf{u}\cdot\hat{\mathbf{b}}, \qquad \frac{\mathrm{d}}{\mathrm{d}t'} = \frac{\mathrm{d}}{\mathrm{d}t}, \tag{29}$$

$$\mu' = \mu, \qquad \alpha' = \alpha, \tag{30}$$

$$\phi' = \phi - \mathbf{u}\cdot\mathbf{A}, \qquad \mathbf{A}' = \mathbf{A}. \tag{31}$$

From these definitions, the transformations for the partial derivatives can be derived:

$$\frac{\partial}{\partial t'} = \frac{\partial}{\partial t} + \mathbf{u}\cdot\nabla + \mathbf{u}\cdot\left(\frac{\partial\hat{\mathbf{b}}}{\partial t} + \mathbf{u}\cdot\nabla\hat{\mathbf{b}}\right)\frac{\partial}{\partial v_\parallel}, \tag{32}$$

$$\nabla' = \nabla + \nabla(\mathbf{u}\cdot\hat{\mathbf{b}})\frac{\partial}{\partial v_\parallel}, \tag{33}$$

$$\frac{\partial}{\partial v_\parallel'} = \frac{\partial}{\partial v_\parallel}, \qquad \frac{\partial}{\partial \mu'} = \frac{\partial}{\partial \mu}, \qquad \frac{\partial}{\partial \alpha'} = \frac{\partial}{\partial \alpha}, \tag{34}$$

leading further to

$$\mathbf{E}' = \mathbf{E} + \mathbf{u}\times\mathbf{B}, \qquad \mathbf{B}' = \mathbf{B}, \qquad \mathbf{U}_0' = \mathbf{U}_0 - \mathbf{u}, \tag{35}$$

$$\frac{\mathrm{D}}{\mathrm{D}t'} = \frac{\mathrm{D}}{\mathrm{D}t} + \left(\mathbf{u}\cdot\frac{\partial\hat{\mathbf{b}}}{\partial t} + \mathbf{U}_0\cdot\nabla(\mathbf{u}\cdot\hat{\mathbf{b}})\right)\frac{\partial}{\partial v_\parallel}. \tag{36}$$

To prove the Galilean covariance of Eq. (23), it suffices to show that $\mathbf{U}_1$ is Galilean invariant since $\dot{\mathbf{X}}$ and $\mathbf{U}_0$ transform like velocity vectors. To show $B_\parallel^*$ is Galilean invariant, we write

$$\begin{aligned}\hat{\mathbf{b}}'\cdot(\nabla'\times\mathbf{U}_0') &= \hat{\mathbf{b}}\cdot(\nabla\times\mathbf{U}_0) + \hat{\mathbf{b}}\cdot\left(\nabla(\mathbf{u}\cdot\hat{\mathbf{b}})\times\frac{\partial\mathbf{U}_0}{\partial v_\parallel}\right)\\ &= \hat{\mathbf{b}}\cdot(\nabla\times\mathbf{U}_0) + \hat{\mathbf{b}}\cdot\left(\nabla(\mathbf{u}\cdot\hat{\mathbf{b}})\times\hat{\mathbf{b}}\right)\\ &= \hat{\mathbf{b}}\cdot(\nabla\times\mathbf{U}_0).\end{aligned} \tag{37}$$

Meanwhile,

$$\begin{aligned}\frac{\mathrm{D}\mathbf{U}_0'}{\mathrm{D}t'} &= \frac{\mathrm{D}\mathbf{U}_0}{\mathrm{D}t} + \left(\mathbf{u}\cdot\frac{\partial\hat{\mathbf{b}}}{\partial t} + \mathbf{U}_0\cdot\nabla(\mathbf{u}\cdot\hat{\mathbf{b}})\right)\frac{\partial\mathbf{U}_0}{\partial v_\parallel}\\ &= \frac{\mathrm{D}\mathbf{U}_0}{\mathrm{D}t} + \left(\mathbf{u}\cdot\frac{\partial\hat{\mathbf{b}}}{\partial t} + \mathbf{U}_0\cdot\nabla(\mathbf{u}\cdot\hat{\mathbf{b}})\right)\hat{\mathbf{b}}.\end{aligned} \tag{38}$$

Since the additional term is parallel to the magnetic field, we find that

$$\hat{\mathbf{b}}'\times\frac{\mathrm{D}\mathbf{U}_0'}{\mathrm{D}t'} = \hat{\mathbf{b}}\times\frac{\mathrm{D}\mathbf{U}_0}{\mathrm{D}t}, \tag{39}$$

leading to

$$\mathbf{U}_1' = \mathbf{U}_1. \tag{40}$$

Therefore,

$$\frac{\mathrm{d}\mathbf{X}}{\mathrm{d}t} = \mathbf{U}_0 + \mathbf{U}_1 \Rightarrow \frac{\mathrm{d}\mathbf{X}'}{\mathrm{d}t'} = \mathbf{U}_0' + \mathbf{U}_1'. \tag{41}$$

Proving that Eq. (24) is Galilean covariant is also straightforward. The parallel electric force and the mirror force are clearly Galilean invariants, and so is $\dot{\mathbf{U}}_0$. The only remaining term to check is the last term:

$$\begin{aligned}\mathbf{U}_1'\cdot\left(\hat{\mathbf{b}}'\cdot\nabla\mathbf{U}_0'\right) &= \mathbf{U}_1\cdot\left(\hat{\mathbf{b}}\cdot\nabla\mathbf{U}_0 + \hat{\mathbf{b}}\cdot\hat{\mathbf{b}}\cdot\nabla(\mathbf{u}\cdot\hat{\mathbf{b}})\frac{\partial\mathbf{U}_0}{\partial v_\parallel}\right)\\ &= \mathbf{U}_1\cdot\left(\hat{\mathbf{b}}\cdot\nabla\mathbf{U}_0 + \hat{\mathbf{b}}\hat{\mathbf{b}}\cdot\nabla(\mathbf{u}\cdot\hat{\mathbf{b}})\right)\\ &= \mathbf{U}_1\cdot\left(\hat{\mathbf{b}}\cdot\nabla\mathbf{U}_0\right).\end{aligned} \tag{42}$$

Therefore,

$$\begin{gathered}\hat{\mathbf{b}}\cdot\frac{\mathrm{d}\mathbf{U}_0}{\mathrm{d}t} = \frac{e}{m}\hat{\mathbf{b}}\cdot\mathbf{E} - \frac{\mu}{m}\hat{\mathbf{b}}\cdot\nabla B + \mathbf{U}_1\cdot\left(\hat{\mathbf{b}}\cdot\nabla\mathbf{U}_0\right),\\ \Downarrow\\ \hat{\mathbf{b}}'\cdot\frac{\mathrm{d}\mathbf{U}_0'}{\mathrm{d}t'} = \frac{e}{m}\hat{\mathbf{b}}'\cdot\mathbf{E}' - \frac{\mu'}{m}\hat{\mathbf{b}}'\cdot\nabla' B' + \mathbf{U}_1'\cdot\left(\hat{\mathbf{b}}'\cdot\nabla'\mathbf{U}_0'\right).\end{gathered} \tag{43}$$

This completes the proof.

## III. GALILEAN INVARIANCE OF SECOND-ORDER GUIDING CENTER QUASINEUTRAL VLASOV THEORY

As we saw in Sec. II, proving that the equations of motion are Galilean covariant requires many manual calculations. This can become quite tedious for higher-order theories. However, if the Lagrangian differs only by a total derivative under a Galilean boost, then Galilean boosts are a divergence symmetry of the action. This implies that the equations of motion are Galilean covariant, and in practice, it is more efficient to prove Galilean invariance from the Lagrangian. In this section, we demonstrate this with a second-order guiding center Lagrangian.[20] The Lagrangian and Hamiltonian are

$$L = \left(e\mathbf{A} + mv_\parallel\hat{\mathbf{b}} + m\mathbf{u}_E - \frac{m\mu}{e}\overline{\mathbf{R}}\right)\cdot\frac{\mathrm{d}\mathbf{X}}{\mathrm{d}t} + \frac{m\mu}{e}\frac{\mathrm{d}\alpha}{\mathrm{d}t} - H, \tag{44}$$

$$H = e\phi + \frac{m}{2}\left(v_\parallel\hat{\mathbf{b}} + \mathbf{u}_E\right)^2 - \frac{m\mu}{2e}\nabla\cdot\left(\hat{\mathbf{b}}\times\mathbf{u}_E\right) + \frac{m\mu}{e}S, \tag{45}$$

where

$$\overline{\mathbf{R}} = (\nabla\hat{\perp})\cdot\hat{\boldsymbol{\alpha}} + \frac{1}{2}\nabla\times\hat{\mathbf{b}}, \tag{46}$$

$$S = \hat{\boldsymbol{\alpha}}\cdot\frac{\partial\hat{\perp}}{\partial t}. \tag{47}$$

The additional terms arise from the second-order analysis and are associated with gyrogauge invariance. Under a Galilean boost, they transform as

$$\overline{\mathbf{R}}' = \overline{\mathbf{R}}, \qquad S' = S + \hat{\boldsymbol{\alpha}}\cdot(\mathbf{u}\cdot\nabla\hat{\perp}). \tag{48}$$

Examining the zeroth and first-order terms first, we find that

$$\mathbf{A}'\cdot\frac{d\mathbf{X}'}{dt'}-\phi'=\mathbf{A}\cdot\frac{d\mathbf{X}}{dt}-\phi, \tag{49}$$

$$\left(v'_\parallel\hat{\mathbf{b}}'+\mathbf{u}'_E\right)\cdot\frac{d\mathbf{X}'}{dt'}-\frac{1}{2}\left(v'_\parallel\hat{\mathbf{b}}'+\mathbf{u}'_E\right)^2 = \left(v_\parallel\hat{\mathbf{b}}+\mathbf{u}_E\right)\cdot\frac{d\mathbf{X}}{dt}-\frac{1}{2}\left(v_\parallel\hat{\mathbf{b}}+\mathbf{u}_E\right)^2+\frac{u^2}{2}-\mathbf{u}\cdot\frac{d\mathbf{X}}{dt}. \tag{50}$$

Meanwhile, the term proportional to $\dot{\alpha}$ is trivial.

For the second-order terms, we require

$$\nabla'\cdot\left(\hat{\mathbf{b}}'\times\mathbf{u}'_E\right)=\nabla\cdot\left(\hat{\mathbf{b}}\times\mathbf{u}_E\right)-\nabla\cdot\left(\hat{\mathbf{b}}\times\mathbf{u}\right) = \nabla\cdot\left(\hat{\mathbf{b}}\times\mathbf{u}_E\right)-\mathbf{u}\cdot\left(\nabla\times\hat{\mathbf{b}}\right). \tag{51}$$

We then obtain

$$-\overline{\mathbf{R}}'\cdot\frac{d\mathbf{X}'}{dt'}+\frac{1}{2}\nabla'\cdot\left(\hat{\mathbf{b}}'\times\mathbf{u}'_E\right)-S' = -\overline{\mathbf{R}}\cdot\frac{d\mathbf{X}}{dt}-\frac{1}{2}\hat{\mathbf{b}}\cdot(\nabla\times\mathbf{u}_E)-S+\Delta, \tag{52}$$

where

$$\Delta=\overline{\mathbf{R}}\cdot\mathbf{u}-\frac{1}{2}\mathbf{u}\cdot\left(\nabla\times\hat{\mathbf{b}}\right)-\hat{\boldsymbol{\alpha}}\cdot(\mathbf{u}\cdot\nabla)=0. \tag{53}$$

Remarkably, the second-order change is zero. The Lagrangian, therefore, transforms as

$$L'=L+\frac{mu^2}{2}-\mathbf{u}\cdot\frac{d\mathbf{X}}{dt}=L+\frac{d}{dt}\left(\frac{mu^2}{2}t-m\mathbf{u}\cdot\mathbf{X}\right). \tag{54}$$

The second-order guiding center equations of motion in the strong flow regime are therefore Galilean covariant.

We now investigate the second-order guiding center QV theory. Let $\mathbf{z}$ denotes the total single particle phase space. We also define a reference phase space $\mathbf{z}_{0s}$ for each species and a reference distribution function $f_{0s}$. The total Lagrangian is composed of the particle Lagrangian and the field Lagrangian:

$$L=\sum_s\int d^6z_{0s}f_{0s}(\mathbf{z}_{0s})\mathcal{L}_{ps}-\frac{1}{2\mu_0}\int d^3x|\nabla\times\mathbf{A}|^2, \tag{55}$$

where $\mathcal{L}_{ps}$ is the single-particle guiding center Lagrangian for species $s$ evaluated at the reference phase space $\mathbf{z}_{0s}$. Since Galilean boosts are a divergence symmetry of the single particle action and the field part of the Lagrangian is unchanged by a Galilean boost, they are a divergence symmetry of the total action. After varying the action, we obtain the second-order guiding center Vlasov equation

$$\frac{\partial f_s}{\partial t}+\dot{\mathbf{X}}\cdot\nabla f_s+\dot{v}_\parallel\frac{\partial f_s}{\partial v_\parallel}=0, \tag{56}$$

and the field equations

$$\nabla\cdot\mathbf{P}=\rho_{gc}, \tag{57}$$

$$\nabla\cdot\mathbf{B}=0, \tag{58}$$

$$\nabla\times\mathbf{E}=-\frac{\partial\mathbf{B}}{\partial t}, \tag{59}$$

$$\nabla\times\mathbf{B}=\mu_0\mathbf{J}_{gc}+\mu_0\frac{\partial\mathbf{P}}{\partial t}+\mu_0\nabla\times\mathbf{M}. \tag{60}$$

Here, $\rho_{gc}$ and $\mathbf{J}_{gc}$ are the guiding center charge density and current and $\mathbf{P}$ and $\mathbf{M}$ are the polarization and magnetization fields, respectively. We refer the reader to Refs. 19–21 for more complete details and definitions. The point is that the resulting QV theory is Galilean invariant because the single particle theory is Galilean invariant, and the electromagnetic contribution respects Galilean invariance. It is straightforward to verify that the new fields transform under Galilean boosts as

$$\rho'_{gc}=\rho_{gc}, \tag{61}$$

$$\mathbf{J}'_{gc}=\mathbf{J}_{gc}-\rho_{gc}\mathbf{u}, \tag{62}$$

$$\mathbf{P}'=\mathbf{P}, \tag{63}$$

$$\mathbf{M}'=\mathbf{M}+\mathbf{u}\times\mathbf{P}. \tag{64}$$

These transformations guarantee Galilean covariance of the field equations and are analogous to the Galilean transformations that arise when analyzing Maxwell's equations in matter.[3] Thus, the second-order guiding center QV theory is Galilean invariant.

## IV. GALILEAN INVARIANCE TO ALL ORDERS

So far, we have demonstrated the Galilean invariance for Hamiltonian guiding center theory up to second order. This prompts the following question: can it be made Galilean invariant to all orders? We prove that the answer is yes. We refer the reader to Ref. 22 for further details of the mathematical machinery employed in this section.

First, we write the equations of motion for a single charged particle. We introduce the orbit parameter $\tau$ to enlarge the phase space to $(\mathbf{x},\mathbf{v},t)$; this will prove convenient, as the Galilean transformation and the background electromagnetic fields are time dependent. After normalizing all relevant quantities and using the ordering parameter $\epsilon=\rho/L$, where $\rho$ is the gyroradius and $L$ is the physical length scale of the system, the equations of motion are

$$\frac{d\mathbf{v}}{d\tau}=\mathbf{E}_0+\mathbf{v}\times\mathbf{B}_0+\epsilon(\mathbf{E}_1+\mathbf{v}\times\mathbf{B}_1), \tag{65}$$

$$\frac{d\mathbf{x}}{d\tau}=\epsilon\mathbf{v}, \tag{66}$$

$$\frac{dt}{d\tau}=\epsilon. \tag{67}$$

We have split up the background electromagnetic fields such that $\mathbf{E}_0\cdot\mathbf{B}_0=0$. It is interesting to note that this requirement is both Galilean invariant and Lorentz invariant, since $\mathbf{E}\cdot\mathbf{B}$ is a frame-invariant scalar in both cases. Notice that in the limit that $\epsilon\to 0$, the motion is purely periodic, and the equations of motion are gyrosymmetric. The corresponding Hamiltonian vector field for this system is

$$X_H=X_0+\epsilon X_1, \tag{68}$$

where

$$X_0=(\mathbf{E}_0+\mathbf{v}\times\mathbf{B}_0)\cdot\frac{\partial}{\partial\mathbf{v}}, \tag{69}$$

$$X_1=(\mathbf{E}_1+\mathbf{v}\times\mathbf{B}_1)\cdot\frac{\partial}{\partial\mathbf{v}}+\mathbf{v}\cdot\frac{\partial}{\partial\mathbf{x}}+\frac{\partial}{\partial t}. \tag{70}$$

Here, $X_0$ is gyrosymmetric while both $X_0$ and $X_1$ are Galilean invariant. This will serve as our base case in the inductive argument to follow.

The goal of guiding center theory is to find an $\epsilon$-dependent near-identity coordinate transformation $T_\epsilon$ such that the equations of motion are gyrosymmetric up to all orders. Using the Lie transform approach, $T_\epsilon$ can be written as

$$T_\epsilon = \ldots \circ \exp(-G_n(\epsilon)) \circ \ldots \circ \exp(-G_1(\epsilon)), \quad (71)$$

where each $G_n$ is a vector field that tends to zero as $\epsilon \to 0$, with each doing so more rapidly than the last. Observe that if each $G_n$ is Galilean invariant, then the coordinate transformation will preserve Galilean symmetry up to all orders, and the modified equations of motion will be Galilean invariant. Thus, the problem reduces to finding a Galilean invariant $G_n$ compatible with the gyrosymmetry requirements of guiding center theory for each $n$. We prove by induction that this can always be done.

First, some preliminaries. The following is a generalization of the tools presented in Ref. 22. Define the mapping $\Phi_\psi$ on the Cartesian phase space coordinates as follows:

$$\begin{aligned}\Phi_\psi(t, \mathbf{x}, \mathbf{v}) = (t, \mathbf{x}, (\mathbf{v}\cdot\hat{\mathbf{b}})\hat{\mathbf{b}} + \mathbf{u}_E \\ + \cos(\psi)\hat{\mathbf{b}} \times \left((\mathbf{v} - \mathbf{u}_E)\times\hat{\mathbf{b}}\right) \\ + \sin(\psi)(\mathbf{v} - \mathbf{u}_E)\times\hat{\mathbf{b}}),\end{aligned} \quad (72)$$

where $\hat{\mathbf{b}}$ and $\mathbf{u}_E$ are defined as usual using $\mathbf{E}_0$ and $\mathbf{B}_0$ as the reference electromagnetic fields. For an aribtrary tensor $\tau$, denote $\tau_\psi = \Phi_\psi^* \tau$, where the asterisk denotes the pullback. We say, a tensor $\tau$ is gyrosymmetric if $\tau_\psi = \tau$ for all real numbers $\psi$. Moreover, $\tau_\psi$ has the following Fourier expansion:

$$\tau_\psi = \langle\tau\rangle + \sum_{k=1}^{\infty} (\Pi_k \tau)\cos(\psi) + (\overline{\Pi}_k \tau)\sin(\psi), \quad (73)$$

$\langle\ldots\rangle$ denotes the gyroaverage, and the Fourier coefficients are defined as usual. Let $\mathcal{L}_Y$ denotes the Lie derivative with respect to $Y$, and defines

$$\xi = \left(\frac{\mathbf{E}_0}{B_0} + \mathbf{v}\times\hat{\mathbf{b}}_0\right)\cdot\frac{\partial}{\partial\mathbf{v}}. \quad (74)$$

Then, the differential operator $\mathcal{L}_\xi$ is an algebraic operator on gyroharmonics, meaning that it can be easily inverted. It is also a Galilean invariant operator. This will prove useful when examining the base case.

Next, the main argument. Let $\mathcal{X}_\epsilon$ be a vector field of the form

$$\mathcal{X}_\epsilon = X_0 + \epsilon X_1 + \cdots + \epsilon_N X_N + \sum_{k=1}^{\infty} \epsilon^{N+k}\chi_k, \quad (75)$$

such that every $X_n$ is both gyrosymmetric and Galiliean invariant, $X_0 = B_0\,\xi$, every $\chi_n$ is Galilean invariant, and $N \geq 0$. The Hamiltonian vector field $X_H$ is clearly of this form with $N = 0$. We claim it is possible to find a transformation $\exp(-\epsilon^{N+1} G_{N+1})$ for some $\epsilon$-independent vector field $G_{N+1}$ such that the transformed $\mathcal{X}_\epsilon$ will be of the same form, but with $N$ replaced with $N+1$. That is, the vector field will be gyrosymmetric to one higher order, and all individual terms in the vector field will still be Galilean invariant. If we can establish our claim, the principle of mathematical induction implies that there is an infinite sequence of near-identity transformations that transforms $\mathcal{X}_\epsilon$ into a nearby vector field that is both gyrosymmetric and Galilean invariant to all orders.

To establish our claim, first denote the transformed vector field as $\mathcal{X}'_\epsilon = \exp(-\epsilon^{N+1}G_{N+1})_*\mathcal{X}_\epsilon$. Expand this as a series of Lie derivatives and observe that the first $N+1$ terms in the series agree with $\mathcal{X}_\epsilon$, while the $(N+2)^{\mathrm{nd}}$ term is

$$\mathcal{X}'_{N+1} = \mathcal{L}_{G_{N+1}} X_0 + \chi_1.$$

This term will be gyrosymmetric if and only if its fluctuating part, $\widetilde{\mathcal{X}}'_{N+1} = \mathcal{X}'_{N+1} - \langle\mathcal{X}'_{N+1}\rangle$, vanishes. The fluctuating part will vanish if and only if the vector field $G_{N+1}$ satisfies the differential equation

$$\mathcal{L}_{X_0}\widetilde{G}_{N+1} = \widetilde{\chi}_1. \quad (76)$$

The unique solution of this equation with zero gyroaverage is

$$G_{N+1} = I\left(\frac{\widetilde{\chi}_1}{B_0}\right) + I^2\left(\mathcal{L}_{\widetilde{\chi}_1/B_0}\ln B_0\right)\xi, \quad (77)$$

where $I$ denotes the inverse of $\mathcal{L}_\xi$. Since the operator $\mathcal{L}_\xi$ is Galilean invariant, the inverse $I$ is also Galilean invariant. Since $B_0$ is Galilean invariant and $\chi_1$ is Galilean invariant by hypothesis, it follows that $G_{N+1}$ is Galilean invariant. Since $G_{N+1}$ and all of the $\chi_k$ are Galilean invariant each of the terms in $\mathcal{X}'_\epsilon$ that are $O(\epsilon^{N+2})$ and higher will also be Galilean invariant. This establishes our claim.

As a final note, the Galilean invariance of the single-particle theory implies Galilean invariance of the QV theory, as discussed in Sec. III. Thus, one can construct a Hamiltonian guiding center QV theory that is Galilean invariant to all orders.

## V. DISCUSSION

Although guiding center theory can be made Galilean invariant, not all guiding center theories are so in practice. We examine two reasons why Galilean invariance can fail.

First, many guiding center theories operate in the weak flow regime, where the $E$-cross-$B$ velocity is small compared to the typical thermal velocity. An example of a first-order Lagrangian is

$$L = \left(e\mathbf{A} + mv_\parallel\hat{\mathbf{b}}\right)\cdot\frac{\mathrm{d}\mathbf{X}}{\mathrm{d}t} + \frac{m\mu}{e}\frac{\mathrm{d}\alpha}{\mathrm{d}t} - H, \quad (78)$$

$$H = \frac{mv_\parallel^2}{2} + \mu B + e\phi. \quad (79)$$

It is easy to verify that the equations of motion that arise from this theory are not Galilean covariant. This arises because the initial assumption of the model, i.e., weak flow, does not respect Galilean symmetry. The perpendicular part of the electric field can be made arbitrarily large via a Galilean boost, so there always exists an inertial reference frame where the theory breaks down. Moreover, the lowest order magnetic moment in the weak flow regime, $\mu = mv_\perp^2/2B$, is not Galilean invariant, meaning that $\dot{\mu} = 0$ is not a Galilean covariant equation. In contrast, the lowest order magnetic moment in the strong flow regime, $\mu = m(\mathbf{v}_\perp - \mathbf{u}_E)^2/2B$, is indeed a Galilean invariant. So, a Galilean invariant guiding center theory must allow for strong flows.

Second, in Sec. IV, we set $\langle G\rangle = 0$ since it was arbitrary. Nonzero choices of $\langle G\rangle$, however, can spoil Galilean symmetry if $\langle G\rangle$ is not Galilean invariant. An example of this can be found in Ref. 7, where the strong flow first-order Lagrangian described in the main text is

$$L = \left(e\mathbf{A} + mv_\parallel \hat{\mathbf{b}}\right) \cdot \frac{\mathrm{d}\mathbf{X}}{\mathrm{d}t} + \frac{m\mu}{e}\frac{\mathrm{d}\alpha}{\mathrm{d}t} - H, \tag{80}$$

$$H = e\phi + \frac{m}{2}\left(v_\parallel \hat{\mathbf{b}} - \mathbf{u}_E\right)^2 + \mu B. \tag{81}$$

The theory is not Galilean invariant. This is because the displacement vector chosen in Eq. (3.5) of Ref. 7 is not Galilean invariant since it is proportional to $\mathbf{E}_\perp$. This is equivalent to choosing $\langle G_1 \rangle$ such that it is not Galilean invariant. In contrast, the Lagrangian found in the appendix of Ref. 7 (dubbed the Northrop Lagrangian) respects Galilean symmetry.

## VI. CONCLUSIONS

In this article, we proved that one can preserve the Galilean invariance of Hamiltonian guiding center theory to all orders, ensuring that the guiding center equations are valid in all inertial reference frames. We demonstrated the Galilean invariance of a first-order theory by analyzing the equations of motion directly and of a second-order theory by analyzing the Lagrangian. We also showed that the resulting QV theory will be Galilean invariant and gave transformation laws for the resulting guiding center fields. One limitation of this work is that because the formulation is Hamiltonian, the kinetic equation is collisionless. An interesting avenue for future work would be analyzing guiding center kinetic theory that includes Galilean invariant collision operators, such as those described in Refs. 23 and 24.

## ACKNOWLEDGMENTS

The authors would like to thank Norman Cao and Max Ruth for insightful discussions.

## AUTHOR DECLARATIONS

### Conflict of Interest

The authors have no conflicts to disclose.

### Author Contributions

**C. D. Stephens:** Conceptualization (lead); Formal analysis (equal); Writing – original draft (lead); Writing – review & editing (equal). **J. W. Burby:** Formal analysis (equal); Writing – original draft (supporting); Writing – review & editing (equal).

## DATA AVAILABILITY

Data sharing is not applicable to this article as no new data were created or analyzed in this study.